\documentclass
[twocolumn,aps,prd,amsmath,showpacs,amssymb,floatfix,nofootinbib]
{revtex4-1}

\usepackage{CJK}       
\usepackage{amssymb}
\usepackage{mathptmx}                % math+times fonts
\usepackage{graphicx}
\usepackage{multirow}
\usepackage{bm}
\usepackage{array}
\usepackage{xcolor}

\def\bea{\begin{eqnarray}}
\def\eea{\end{eqnarray}}
\def\be{\begin{equation}}
\def\ee{\end{equation}}
\def\non{\nonumber}

\def\al{\alpha}
\def\eps{\varepsilon}
\def\ms{M_\odot}

\begin{document}

\title{Rotating hybrid stars with the Dyson-Schwinger quark model}

\begin{CJK*}{GB}{gbsn}
\author{J.-B. Wei (κ½ð±ê)}
\author{H. Chen (³Â»¶)}
\email[]{huanchen@cug.edu.cn}
\affiliation{
School of Mathematics and Physics, China University of Geosciences,
Lumo Road 388, 430074 Wuhan, China}

\author{G. F. Burgio}
\author{H.-J. Schulze}
\affiliation{
INFN Sezione di Catania, Dipartimento di Fisica,
Universit\'a di Catania, Via Santa Sofia 64, 95123 Catania, Italy}

\begin{abstract}
We study rapidly rotating hybrid stars with the
Dyson-Schwinger model for quark matter
and the Brueckner-Hartree-Fock many-body theory with realistic two-body and
three-body forces for nuclear matter.
We determine the maximum gravitational mass, equatorial radius,
and rotation frequency of stable stellar configurations
by considering the constraints of the Keplerian limit and
the secular axisymmetric instability,
and compare with observational data.
We also discuss the rotational evolution for constant baryonic mass,
and find a spinup phenomenon for supramassive stars
before they collapse to black holes.
\end{abstract}

\maketitle

\end{CJK*}

%-------------------------------------------------------------------------------
\section{Introduction}

Neutron stars (NS) are among the densest objects known in the Universe.
They contain an extreme environment shaped by the effects of the four
fundamental interactions.
NSs have the typical mass $M \sim 1.4\ms$ and radius $R\sim10$km.
Therefore, the mean particle density can reach (2--3)$\rho_0$,
and the core density (10--20)$\rho_0$ \cite{Haensel2007},
where $\rho_0=0.17\,\text{fm}^{-3}$ is the so-called nuclear saturation density.
At this density, the nucleons might undergo a phase transition to
quark matter (QM),
and a hybrid NS (HNS) with a QM core is formed.
This makes NS ideal astrophysical laboratories
to study hadronic interactions over a wide range of densities \cite{Weber2014}.

Unfortunately, as a key ingredient of the investigation of NS,
the equation of state (EOS) remains uncertain.
The microscopic theory of the nucleonic EOS has reached a high degree
of sophistication \cite{gle,bbb,akma,zhou,Li2008,mmy},
but the QM EOS is poorly known
at zero temperature and at the high baryonic density appropriate for NS,
because it is difficult to perform first-principle calculations of QM.

Therefore one can presently only resort to more or less phenomenological models
for describing QM,
such as the MIT Bag model \cite{Chod},
the Nambu-Jona-Lasino model \cite{Buballa05,Schertler99,Klahn,Klahn15},
or the quasi-particle model \cite{Tian2012,Zhao2015}.
In Ref.~\cite{Chen2011} we developed a Dyson-Schwinger quark model (DSM)
for deconfined QM,
which provides a continuum approach to QCD that can simultaneously
address both confinement and dynamical chiral symmetry breaking
\cite{Roberts1994,Alkofer2000wg}.
In that work, we considered static and spherical symmetric HNSs,
whereas in this paper we include the effects of rotation.

Rotation is a common property of NS.
Of the thousands of currently observed pulsars,
the fastest one has been discovered in the globular cluster Terzan 5 with a
frequency of 716~Hz \cite{716hz}.
At this rapid rotation, a NS would be flattened by the centrifugal force,
and the Tolman-Oppenheimer-Volkoff equation,
suitable for a static and spherically symmetric situation,
cannot describe correctly the rotating stellar structure.
In the present paper we approximate the NS as a axisymmetric and rigid
rotating body,
and resort to Einstein's theory of general relativity
for a rapidly rotating star.
Numerical methods for (axisymmetric) rotating stellar structure
have been advanced by several groups
\cite{Komatsu1989a,Weber1991a,Weber1991b,Cook1992,Cook1994,Salg1994,
Stergioulas2003,Stergioulas1995}.
In this work we utilize the KEH method \cite{Komatsu1989a} to
obtain the properties of rapidly rotating HNSs.

This paper is organized as follows.
In Sec.~II we briefly discuss
the construction of the EOS of a HNS.
In Sec.~III we present the rotation effects on the HNS;
the allowed ranges of gravitational mass, equatorial radius,
and Kepler frequency are discussed in this section
and compared with observational data.
The rotational evolution for a constant baryonic mass is also analyzed.
Sec.~IV contains our conclusions.

%-------------------------------------------------------------------------------
\section{The equation of state}

\subsection{Nuclear matter}

For nuclear matter
we resort to the Brueckner-Hartree-Fock (BHF) many-body theory
with realistic two-body and three-body nucleonic forces,
which has been extensively discussed in Ref.~\cite{Baldo1999}.
We recall that this theory has also been extended with the inclusion of hyperons,
which might appear in the core of a NS.
The hyperonic EOS in this theory turns out to be very soft,
and this results in too low NS maximum masses \cite{Schulze2011},
%$M<1.7\,\ms$ ($\ms\approx2\times10^{33}$g),
well below the current observational limit
of about two solar masses \cite{heavy,heavy2,heavy3}.
The presence of strange baryonic matter often inhibits
the appearance of QM.
In this work we do not discuss this aspect,
but limit ourselves to consider only nucleons and leptons
in the hadronic phase.

In the BHF theory the energy per nucleon of nuclear matter is given by
\be
 {B \over A}  = {3\over 5}{k^2_F \over 2m}
 + {1\over 2\rho} \sum_{k,k'<k_F}\!\!\!
 \big\langle kk' \big| G[e(k)+e(k');\rho] \big| kk' \big\rangle_A \:,
\ee
where $G[E;\rho]$ is the solution of the Bethe-Goldstone equation
\be
 G[E;\rho] = V + \!\!\!\sum_{k_a,k_b>k_F}\!\!\! V
 {\big| k_a,k_b \big\rangle Q \big\langle k_a,k_b \big| \over E-e(k_a)-e(k_b)}
 G[E;\rho] \:,
\ee
$V$ is the bare nucleon-nucleon (NN) interaction,
$\rho$ is the nucleon number density,
and $E$ the starting energy.
The single-particle energy
\be
 e(k) = e(k;\rho) = {k^2\over 2m} + U(k;\rho) %\:,
\label{e:en}
\ee
and the Pauli operator $Q$ determine the propagation of
intermediate baryon pairs.
The BHF approximation for the
single-particle potential using the {\it continuous choice} is
\be
 U(k;\rho) = \sum _{k'\leq k_F}
 \big\langle k k' \big| G[e(k)+e(k'); \rho] \big| k k' \big\rangle_A \:.
\ee
Due to the occurrence of $U(k)$ in Eq.~(\ref{e:en}),
the above equations constitute a coupled system that has to be solved
in a self-consistent manner for several momenta of the particles involved,
at the considered densities.
The only input quantities of the calculation are the NN
two-body potentials.
In this work we present results obtained with
the Bonn-B (BOB) potential \cite{bob} as input,
supplemented with compatible three-body forces \cite{Li2008,zuotbf,uix}.
The associated EOS yields fairly large maximum masses of about
$2.5\ms$ for purely nucleonic NS (NNS).

For the calculation of the energy per nucleon of asymmetric nuclear matter,
we use the so-called parabolic approximation \cite{bbb}
\be
 {B \over A}(\rho,x) = {B \over A}(\rho,x=0.5)
 + (1-2x)^2 E_\text{sym}(\rho) \:,
\ee
where $x=\rho_p/\rho$ is the proton fraction
and $E_\text{sym}(\rho)$ is the symmetry energy,
which can be expressed in terms of the difference of the energy per nucleon
of pure neutron matter ($x=0$) and symmetric matter ($x=0.5$):
\be
 E_\text{sym}(\rho) = {B \over A}(\rho,x=0) - {B \over A}(\rho,x=0.5) \:.
\ee
The parametrized results of pure neutron and symmetric matter
with different interactions can be found in Ref.~\cite{Li2008}.
The energy density of baryon/lepton matter
as a function of the different partial densities is then
\bea
 \eps(\rho_n,\rho_p,\rho_e,\rho_\mu) &=&
 (\rho_n m_n +\rho_p m_p)
 + (\rho_n+\rho_p) \frac{B}{A}(\rho_n,\rho_p)
\nonumber\\ &&
 + \eps_e(\rho_e) + \eps_\mu(\rho_\mu),
\label{e:epsnn}
\eea
where $\eps_e(\rho_e)$ and $\eps_\mu(\rho_\mu)$ are the energy densities
of electrons and muons.
Once the energy density is known,
the chemical composition of the beta-equilibrated matter can be calculated
and finally the EOS,
\be
 P = \rho^2 {d\over d\rho}
 {\eps(\{\rho_i(\rho)\})\over \rho}
 = \rho {d\eps \over d\rho} - \eps \:.
\ee

%-------------------------------------------------------------------------------
\subsection{Quark matter}

The quark propagator based on the Dyson-Schwinger equation
at finite chemical potential
$\mu \equiv \mu_q = \mu_B/3$
assumes a general form with rotational covariance,
\bea
 S(p;\mu)^{-1} &=&
 i{\bm \gamma}{\bm p} + i \gamma_4 (p_4+i\mu) + m_q + \Sigma(p;\mu)
\\ &\equiv&
 i {\bm \gamma}{\bm p} \;A(p^2,p\cdot u) + B(p^2,p\cdot u)
\nonumber\\&&
 + i \gamma_4(p_4+i\mu) \;C(p^2,p\cdot u) \:,
\label{sinvp}
\eea
where $m_q$ is the current quark mass,
$u=(\bm{0},i\mu)$,
and possibilities of other structures, e.g.,
color superconductivity \cite{Alford2003,Yuan2006,Nickel2006},
are disregarded.
The quark self-energy can be obtained from the gap equation,
\bea
 \Sigma(p;\mu) &=&
 \int\! \frac{d^4q}{(2\pi)^4} \,
 g^2(\mu) D_{\rho\sigma}(p-q;\mu)
\non\\&&\times
 \frac{\lambda^a}{2} \gamma_\rho S(q;\mu)
 \frac{\lambda^a}{2} \Gamma_\sigma(q,p;\mu) \:,
\label{gensigma}
\eea
where $\lambda^a$ are the Gell-Mann matrices,
$g(\mu)$ is the coupling strength,
$D_{\rho\sigma}(k;\mu)$ the dressed gluon propagator,
and $\Gamma_\sigma(q,p;\mu)$ the dressed quark-gluon vertex
at finite chemical potential.

For the quark-gluon vertex and the gluon propagator
we employ the widely-used "rainbow approximation" \cite{Chen2008,Chen2011}
\be
 \Gamma_\sigma(q,p;\mu) = \gamma_\sigma \:,
\ee
and assume the Landau gauge form for the gluon propagator, with an infrared-dominant interaction
modified by the chemical potential \cite{Chen2011,Jiang2013}
\be
 g^2(\mu) D_{\rho \sigma}(k,\mu) =
 4\pi^2 d \frac{k^2}{\omega^6} e^{-\frac{k^2+\al\mu^2}{\omega^2}}
 \Big(\delta_{\rho\sigma}-\frac{k_\rho k_\sigma}{k^2}\Big) \:.
\label{gaussiangluonmu}
\ee
The various parameters can be obtained by fitting meson properties
and chiral condensate in vacuum \cite{Alkofer2002,Chang2009},
and we use
$\omega=0.5\;\text{GeV}$,
$d=1\;\text{GeV}^2$.
The phenomenological parameter $\al$ represents a reduction of the
effective interaction with increasing chemical potential.
This parameter cannot yet be fixed independently
and its value has been amply discussed in previous works
\cite{Chen2011,Chen2015}.

Knowing the quark propagator,
the EOS of cold QM can be obtained via the momentum distribution
\cite{Chen2011,Chen2008,Klahn2009},
\bea
 f_q(|\bm p|;\mu) &=& {1\over4\pi} \int_{-\infty}^{\infty} \!\!\!dp_4\,
 \text{tr}_D\left[-\gamma_4S_q(p;\mu)\right] \:,
\\
 \rho_q(\mu) &=& 6\int \!\!{d^3p\over(2\pi)^3} f_q(|\bm p|;\mu) \:,
\\
 P_q(\mu_q) &=& P_q(\mu_{q,0}) +
 \int_{\mu_{q,0}}^{\mu_q} d\mu \rho_q(\mu) \:.
\eea
The total density and pressure for pure QM are given by summing
the contributions of all flavors.
In addition, we define the phenomenological bag constant
\be
  B_\text{DS} \equiv -\sum_{q=u,d,s} P_q(\mu_{q,0}) \:.
\ee
In this work we set the value as $B_\text{DS}=90\;{\rm MeV\,fm^{-3}}$,
see the discussion in \cite{Chen2011}.

%-------------------------------------------------------------------------------
\subsection{Construction of the hybrid star EOS}

\begin{figure}%.................................................................
\vspace{-4mm}
\centerline{\includegraphics[scale=0.32]{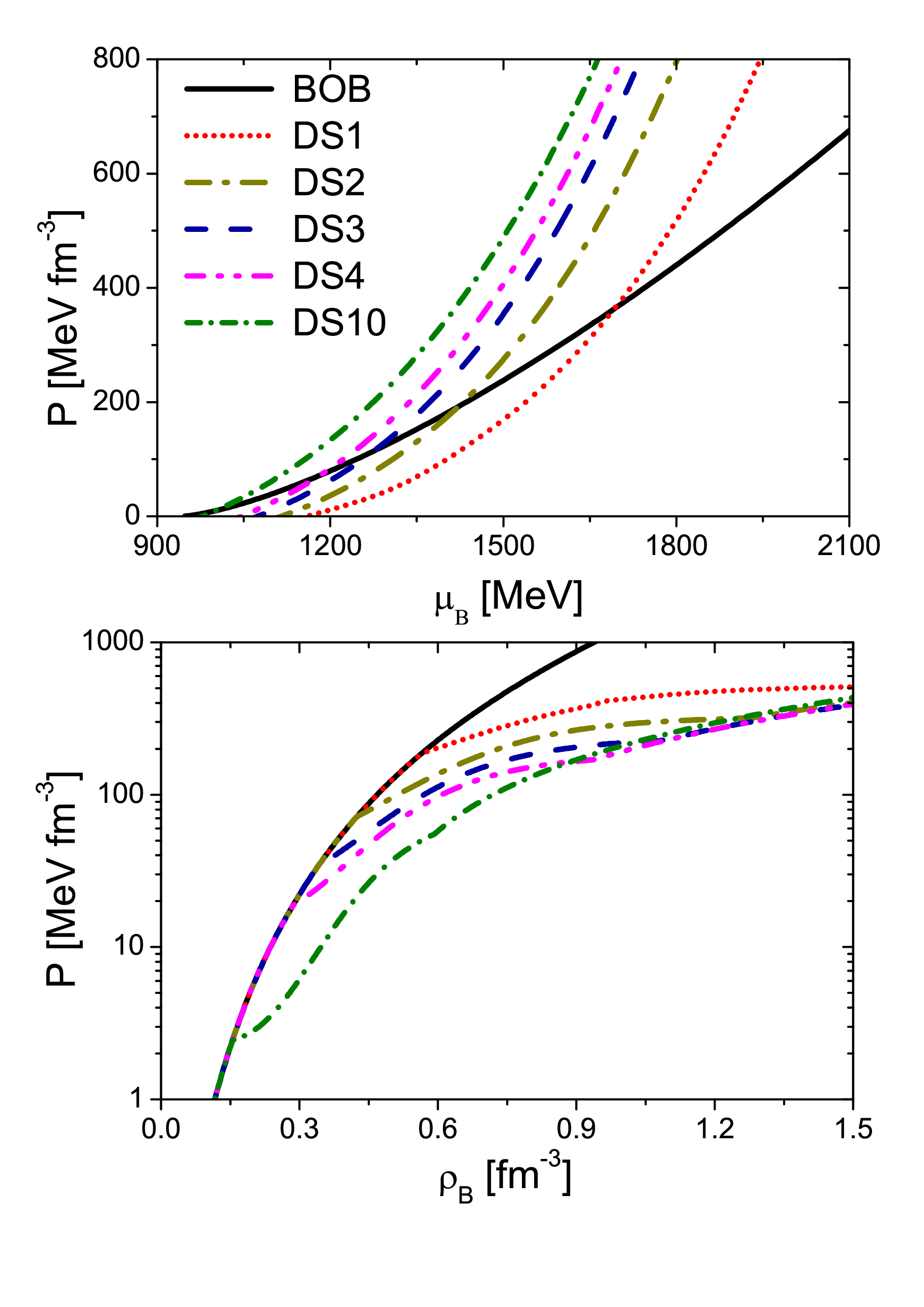}}
\vspace{-12mm}
\caption{(Color online)
Upper panel:
Pressure versus baryon chemical potential
for beta-stable and asymmetric nuclear matter and QM.
The solid curve denotes nuclear matter using the BOB EOS,
and the broken curves labeled DS$\alpha$ represent the DSM EOS
for different choices of~$\alpha$.
Lower panel:
Complete EOS of HNSs with the Gibbs phase transition construction.
}
\label{f:eos}
\end{figure}%...................................................................

In order to study the properties of a rapidly rotating HNS,
we should first construct the EOS of the star.
We assume that the hadron-quark phase transition is of first order,
and perform the Gibbs construction,
thus imposing that nuclear matter and QM are betastable
and globally charge neutral.
This is at variance with the Maxwell construction,
where the two phases must be separately charge neutral.

In the purely nucleonic phase,
which consists of baryons ($n,p$) and leptons ($e,\mu$),
the conditions of  beta stability and charge neutrality can be expressed as
\bea
 && \mu_n-\mu_p = \mu_e = \mu_\mu \:,
\\
 && \rho_p = \rho_e + \rho_\mu \:,
\eea
where $\mu_i$ are the chemical potentials
and $\rho_i$ the particle number densities.
Similarly the pure QM phase,
which contains three-flavor quarks ($u,d,s$) and leptons ($e,\mu$),
should satisfy the constraints of beta stability and charge neutrality
\bea
 && \mu_d = \mu_u+\mu_e = \mu_u+\mu_\mu = \mu_s \:,
\\
 && {2\rho_u-\rho_d-\rho_s\over 3} - \rho_e - \rho_\mu = 0 \:.
\eea

According to the Gibbs construction,
there is a mixed phase where the hadron and quark phases coexist,
and both phases are in equilibrium with each other \cite{gle}.
This can be expressed as
\be
  \mu_i = b_i \mu_B - q_i \mu_e \ , \quad p_H = p_Q = p_M \:.
\ee
where $b_i$ and $q_i$ denote baryon number and charge of the particle species
$i=n,p,u,d,s,e,\mu$ in the mixed phase.
To solve those equations,
we also need the global charge neutrality condition
\be
 \chi\rho_c^Q + (1-\chi)\rho_c^H = 0 \:,
\ee
where $\rho_c^Q$ and $\rho_c^H$ are the charge densities
of quark and nuclear matter,
and $\chi$ is the volume fraction occupied by QM in the mixed phase.
From these equations,
we can derive the energy density $\eps_M$
and the baryon density $\rho_M$ of the mixed phase as
\bea
 \eps_M &=& \chi\eps_Q + (1-\chi)\eps_H \:,
\\
 \rho_M &=& \chi\rho_Q + (1-\chi)\rho_H \:.
\eea

In the upper panel of Fig.~\ref{f:eos}
we show the pressure versus baryon chemical potential
$\mu_B = \mu_n = \mu_u+2\mu_d$.
The solid black curve represents the calculation for
beta-stable and asymmetric nuclear matter with BOB EOS;
the curves labeled DS$\alpha$ are for pure QM
with several choices of the phenomenological parameter $\alpha$.
In the lower panel the complete EOSs of HNSs are shown, i.e.,
pressure vs.~baryon density.
We can see that the EOS contains three sections:
a pure hadronic phase at low density,
followed by a mixed phase,
and a pure quark phase at high density.
We note that the onset of the phase transition is determined by the value
of the parameter $\alpha$;
larger $\alpha$ produces an increasingly softer QM EOS
with a lower phase transition onset density.
For high values of $\alpha$ we find that QM appears quite early, e.g.,
for $\alpha = 10$ at a baryon density $\rho \approx \rho_0$.

For completeness,
we mention that for the calculation of the stellar structure we use the EOSs
by Feynman-Metropolis-Teller \cite{fey} and Baym-Pethick-Sutherland \cite{bps}
for the outer and inner crusts, respectively.

%-------------------------------------------------------------------------------
\section{Results and discussion}

The structure of a rapidly rotating NS is different from the static one,
since the rotation can strongly deform the star.
We assume NS are steadily rotating and have axisymmetric structure.
Therefore the space-time metric used to model a rotating star can be expressed as
\be
  ds^2 = -e^{\gamma+\rho}dt^2
  + e^{2\beta} \left( dr^2+r^2d\theta^2 \right)
  + e^{\gamma-\rho} r^2\sin^2\!\!\theta \left( d\phi-\omega dt \right)^2 \:,
\ee
where the potentials $\gamma,\rho,\beta$,$\omega$ are functions of
$r$ and $\theta$ only.
The matter inside the star is approximated by a perfect fluid
and the energy-momentum tensor is given by
\be
  T^{\mu\nu} = (\eps+p)u^{\mu}u^{\nu} - p g^{\mu\nu} \:,
\ee
where $\eps$, $p$, and $u^{\mu}$ are the energy density,
pressure, and four-velocity, respectively.
In order to solve Einstein's field equation for the potentials
$\gamma,\rho,\beta$,$\omega$,
we adopt the KEH method and use the public RNS code \cite{code}
for calculating the properties of a rotating star.

%-------------------------------------------------------------------------------
\subsection{Keplerian limit} % and secular axisymmetric instability}

\begin{figure}%.................................................................
\vspace{-13mm}
\centerline{\hspace{10mm}\includegraphics[scale=0.37]{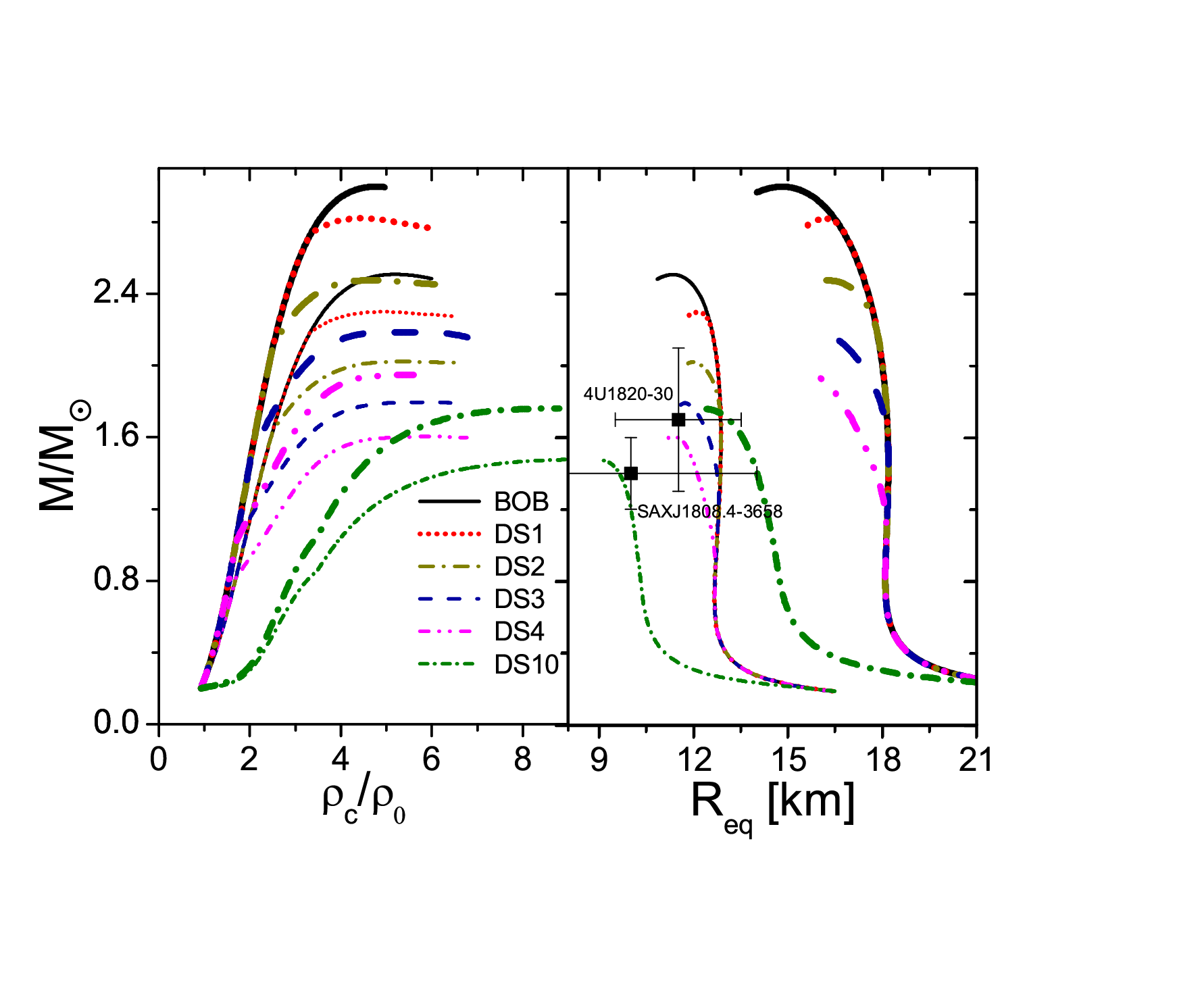}}
\vspace{-19mm}
\caption{(Color online)
Gravitational mass
(in units of the solar mass $\ms=2\times10^{33}$g)
vs.~the normalized
($\rho_0=0.17\;\text{fm}^{-3}$)
central baryon density
(left panel)
and vs.~equatorial radius (right panel)
for different EOSs.
Thin/bold curves denote static/Keplerian sequences.
The observational data are discussed in Sect.~\ref{s:stable}.
}
\label{f:krm}
\end{figure}%...................................................................

\begin{table}%..................................................................
\centering
\caption{
Several properties of rotating NS for the selected EOSs:
Maximum gravitational mass,
corresponding central baryon density,
and Maximum Keplerian frequency.}
\setlength{\tabcolsep}{3pt}
\renewcommand{\arraystretch}{1.2}
\vspace{1mm}
\begin{tabular}{lccccccc}
\hline\hline
\multicolumn{2}{l}{EOS} & BOB  & DS1  & DS2  & DS3  & DS4  & DS10 \\
\hline
\multirow{2}{*}{Static}
  & $M_\text{max}/\ms$  & 2.51 & 2.30 & 2.02 & 1.79 & 1.60 & 1.48 \\
  & $\rho_c/\rho_0$     & 5.22 & 4.96 & 5.38 & 5.88 & 6.14 & 9.69 \\
  \hline
\multirow{3}{*}{Keplerian}
  & $M_\text{max}/\ms$  & 2.99 & 2.82 & 2.47 & 2.19 & 1.95 & 1.76 \\
  & $\rho_c/\rho_0$     & 4.52 & 4.43 & 4.64 & 5.08 & 5.45 & 8.64 \\
  & $f_K\;[\text{Hz}]$  & 1653 & 1461 & 1399 & 1346 & 1316 & 1763 \\
\hline\hline
\end{tabular}
\label{table}
\end{table}%....................................................................

\begin{figure}%.................................................................
\vspace{-8mm}
\centerline{\hspace{20mm}\includegraphics[scale=0.40]{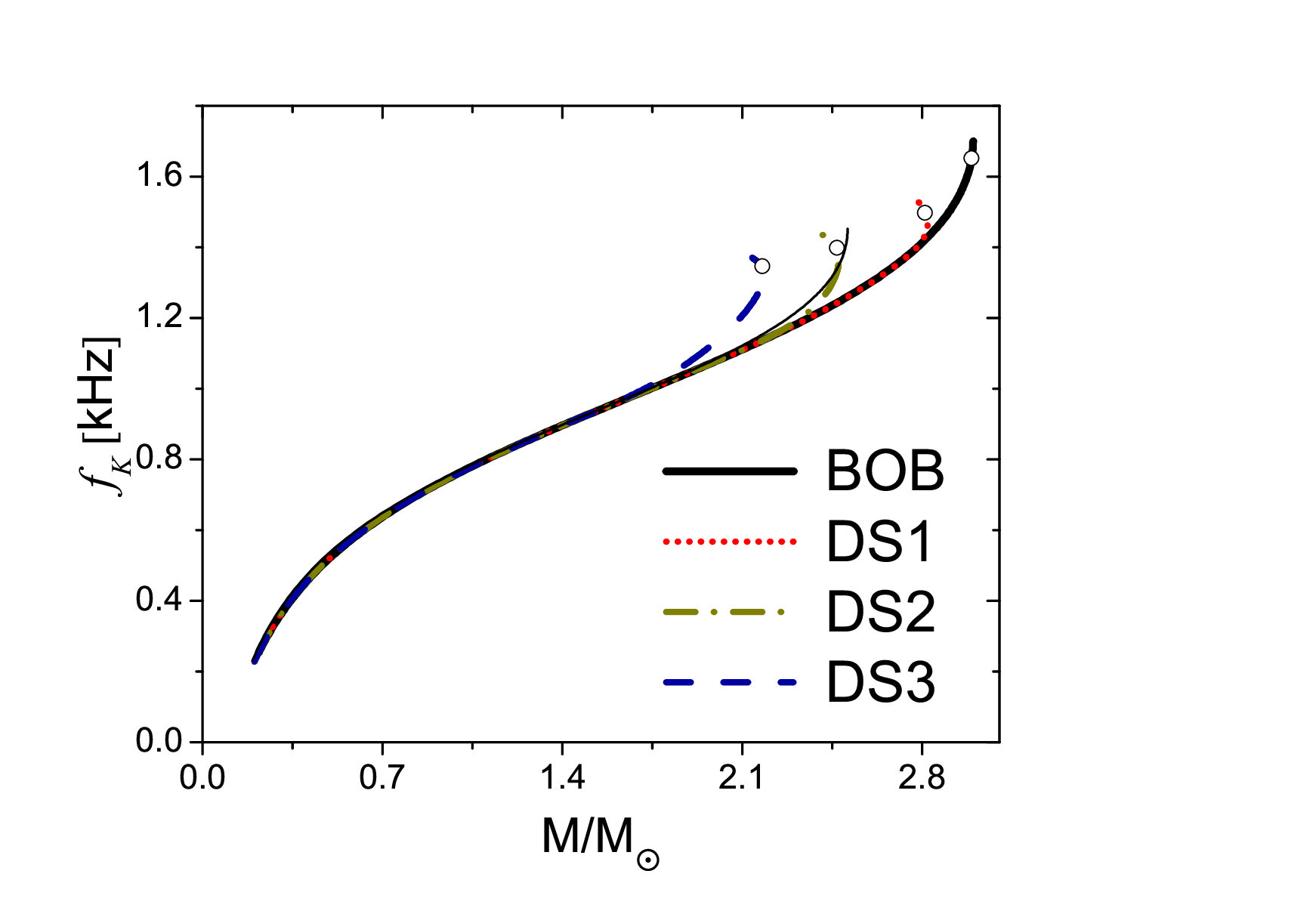}}
\vspace{-7mm}
\caption{(Color online)
Precise (bold curves) and approximated (thin solid curve) values
of Keplerian frequency versus the gravitational mass
for NNSs (BOB) and HNSs (DS$\alpha$).}
\label{f:mk}
\end{figure}%...................................................................

The rotational frequency is a directly measurable quantity of pulsars,
and the Keplerian (mass-shedding) frequency $f_K$
is one of the most-studied physical quantities for rotating stars
\cite{Cook1992,Cook1994,Haensel1995,Benhar2005,Haensel2009,zhang2013}.
In Fig.~\ref{f:krm} we show the gravitational NS mass as a function of the
central baryon density (left panel) and of the equatorial radius (right panel),
using the EOSs displayed in Fig.~\ref{f:eos}.
Results are plotted for both the static configurations (thin curves)
and for the ones rapidly rotating at Keplerian frequency (bold curves).

In all cases
the maximum masses of HNSs are lower than those of NNSs,
because the appearance of QM in the core of the star results in a
softening of the very hard nucleonic EOS.
Comparing Keplerian and static sequences,
rotations increase the maximum mass and equatorial radius substantially.
The maximum masses of the static and Keplerian sequences with various EOSs,
as well as the corresponding central densities,
are listed in Table~\ref{table}.
The maximum masses increase by about $20\%$ from the static to the
Keplerian sequence.
According to the current observations of massive pulsars
\cite{heavy,heavy2,heavy3},
the DSM EOSs with $\alpha\gtrsim2$ are ruled out.

In Fig.~\ref{f:mk} we present the Keplerian frequency
as a function of gravitational mass for some selected EOSs.
We observe that it increases monotonically both for NNSs and HNSs.
%at variance with the behavior of the angular momentum or the gravitational mass
%of stable configurations.
%The maximum Keplerian frequency is obtained from the end curves
%of stable configurations,
%the same as those labelled by open circles in Fig.~\ref{f:jm}.
The Keplerian frequency of HNSs increases more rapidly after QM onset,
and is larger than the one of a NNS with the same gravitational mass,
because the stellar radius is smaller in the former case
due to the presence of a very dense QM core.
However, due to the lower maximum mass of HNSs,
the maximum Keplerian frequency of HNSs is lower than the one of NNSs,
as also listed in Table~\ref{table} for the various EOSs discussed above.
Our results satisfy the constraint from the observed fast-rotating
pulsar PSR J1748-2446ad with $\rm 716\,Hz$ \cite{716hz},
or the even more severe constraint from
XTE J1739-285 with $\rm 1122\,Hz$ \cite{1122hz},
which has not been confirmed, however.

We compare our results with the empirical formula
\be
 f_K = f_0 \left( \frac{M}{\ms} \right)^{\frac{1}{2}}
 \left( \frac{R_s}{10\text{km}} \right)^{-\frac{3}{2}} \:,
\ee
proposed in \cite{Lattimer2004},
where $M$ is the gravitational mass of the Keplerian configuration,
$R_s$ is the radius of the nonrotating configuration of mass $M$,
and $f_0$ is a constant,
which does not depend on the EOS.
In Ref.~\cite{Haensel2009} an optimal prefactor
$f_0=1080\;\text{Hz}$ in the range
$0.5\,\ms < M < 0.9\,M_\text{max}^\text{static}$ was obtained.
Rotating HNSs with masses in that range are characterized
by a purely nucleonic phase,
and therefore the empirical formula cannot be applied.
This is at variance with NNS configurations.
As displayed in Fig.~\ref{f:mk},
our results for NNSs below $2.1\ms$ can be fitted well with
the same parameter $f_0=1080\,\text{Hz}$,
as shown by the thin curve.

%-------------------------------------------------------------------------------
\subsection{Stability analysis}
\label{s:stable}

\begin{figure}[t]%..............................................................
\vspace{-3mm}
\centerline{\includegraphics[scale=0.3]{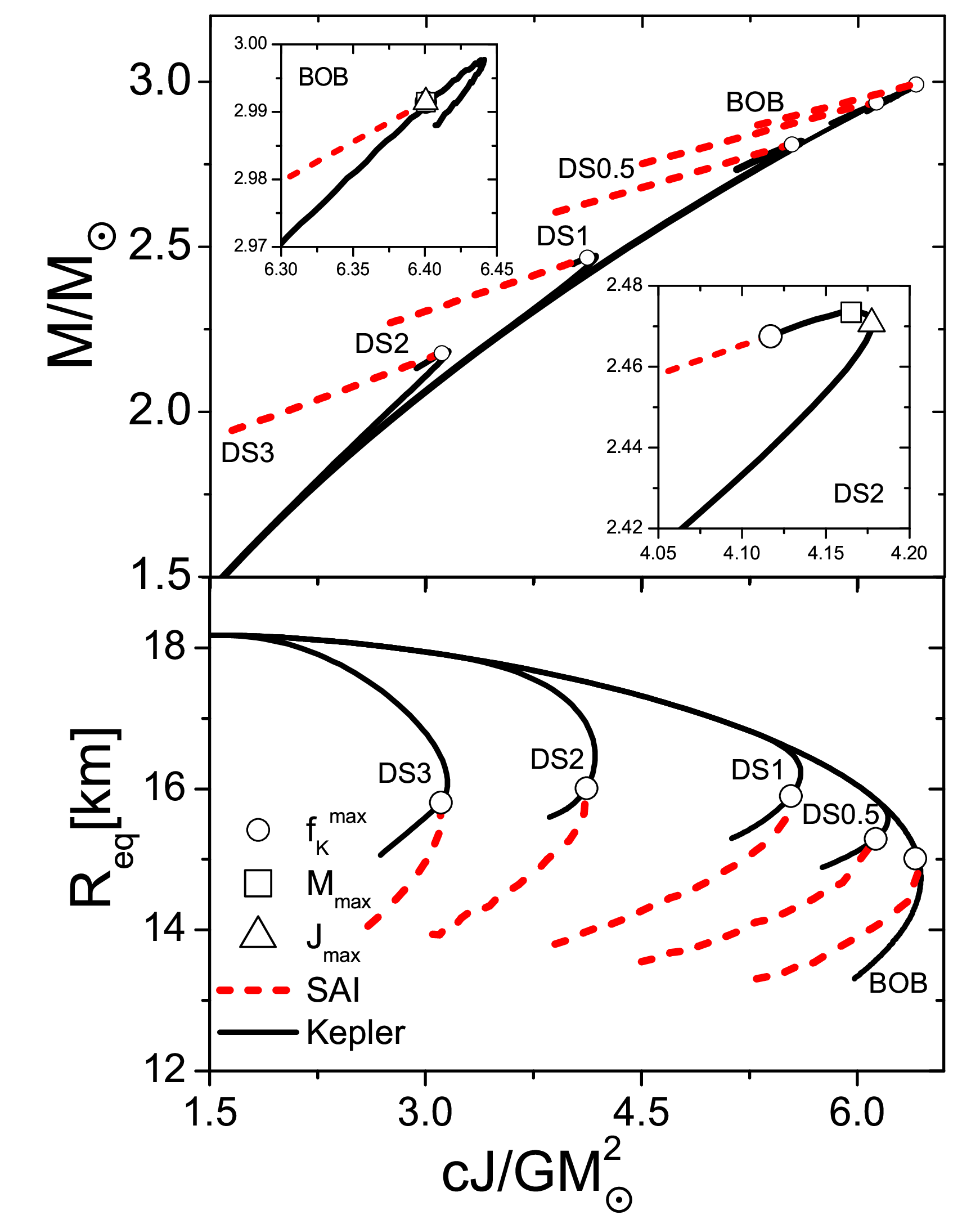}}
\vspace{-4mm}
\caption{(Color online)
Gravitational mass (upper panel)
and radius (lower panel)
versus angular momentum of the Keplerian sequence
(solid black curves)
and SAI
(dashed red curves)
for some selected EOSs.
The open circles represent the SAI onset on the Keplerian sequence.}
\label{f:jm}
\end{figure}%...................................................................

\begin{figure}[t]%..............................................................
\vspace{-14mm}
%\centerline{\hspace{8mm}\includegraphics[scale=0.35]{rho-m}}
%\vspace{-11mm}
\centerline{\hspace{-6mm}\includegraphics[scale=0.32]{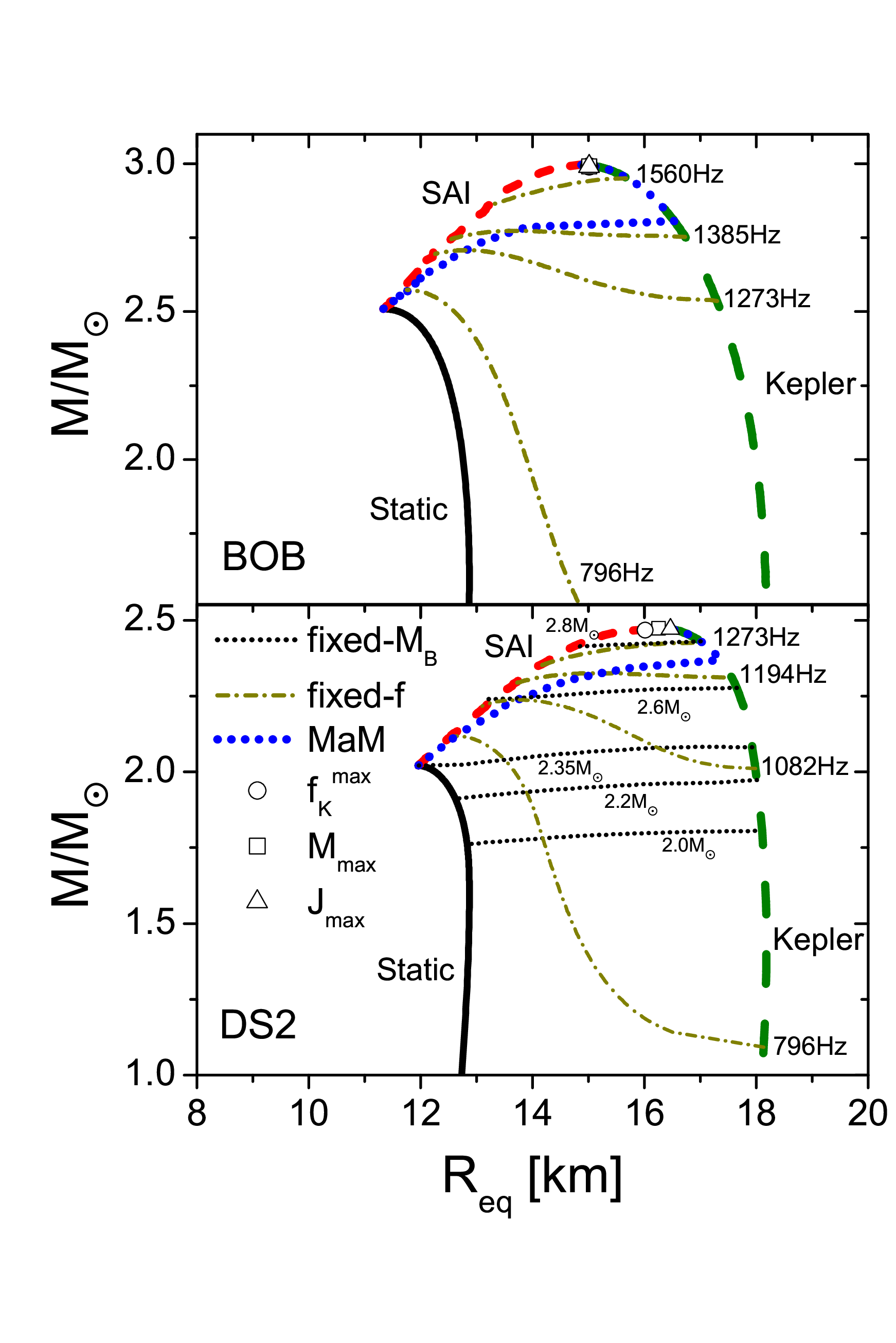}}
\vspace{-16mm}
\caption{(Color online)
%The allowed domain
Mass-radius relations of NS with the EOS BOB (upper panel)
and DS2 (lower panel)
at various fixed rotation frequencies $f$
(dash-dotted olive curves)
or fixed baryonic mass $M_B$
(dotted black curves, discussed with Fig.~\ref{f:fj}).
The positions of the maxima of the fixed-$f$ curves are joined by the
dotted blue curves.}
\label{f:rhomf}
\end{figure}%...................................................................

\begin{figure}%.................................................................
\vspace{-11mm}
\centerline{\includegraphics[scale=0.31]{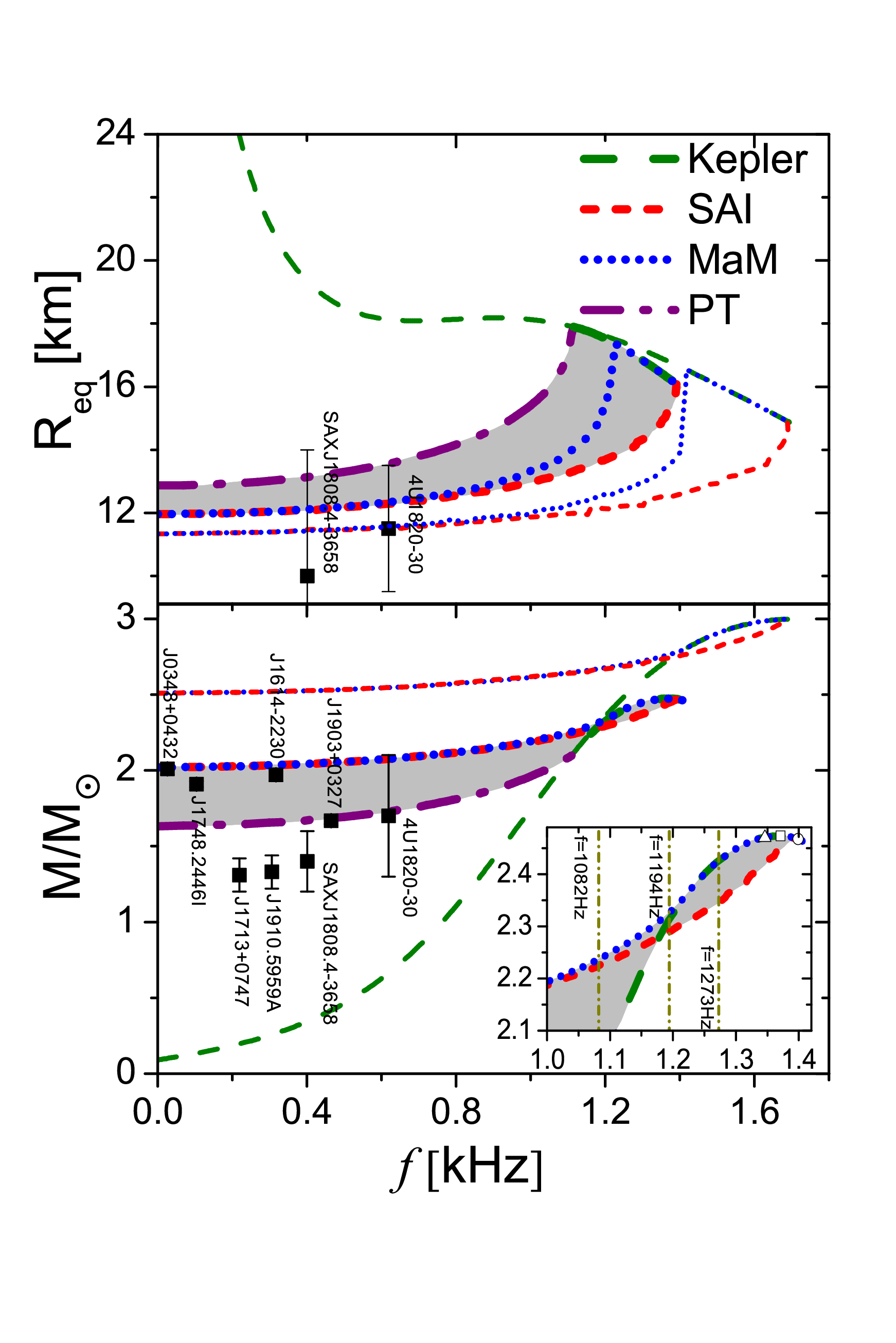}}
\vspace{-17mm}
\caption{(Color online)
The possible values of equatorial radius (upper panel) and
gravitational mass (lower panel)
of NS for the EOS DS2 (bold curves) and BOB (thin curves),
respecting the mass-shedding (dashed green lines)
and SAI limits (short-dashed red lines).
The maximum-mass curves of Fig.~\ref{f:rhomf} are also shown
(MaM, dotted blue lines).
The dash-dotted lila line (PT) indicates the onset of the quark phase
with the DS2 EOS.
The markers represent observational data \cite{Kurkela2010}.}
\label{f:frm}
\end{figure}%...................................................................

In order to complete the description of Figs.~\ref{f:krm} and \ref{f:mk},
one should pay attention to the stability criteria of stars.
It is well known that the onset of the instability of the static sequence
is determined by the condition $dM/d\rho_c = 0$,
i.e., the curve should stop at its mathematical maximum,
which thus gives the maximum mass of the static stable sequence.
In the rotating case, the above criterium has to be generalized, i.e.,
a stellar configuration is stable if its mass $M$ increases with growing
central density for a fixed angular momentum $J$ \cite{Haensel2007}.
Therefore the onset of the instability,
which is called secular axisymmetric instability (SAI),
is expressed by
\be
 \left. \frac{\partial M}{\partial\rho_c}\right|_J = 0 \:.
\label{Kstability}
\ee

The configurations in the Keplerian sequences shown in Fig.~\ref{f:krm}
have different angular momenta,
and thus the curves do not stop at the mathematical maximum.
In the upper panel of Fig.~\ref{f:jm} we show, for some selected EOSs,
the gravitational mass
for the Keplerian sequence vs.~the angular momentum (solid black curves),
along with the SAI condition, Eq.~(\ref{Kstability}),
represented by the dashed red curves.
Thus the Keplerian sequence should stop at the intersection
with the SAI curves, which is indicated by an open circle.
This constraint determines the corresponding endpoints of the curves
in Figs.~\ref{f:krm} and \ref{f:mk}.

Some enlarged details are shown in the insets of Fig.~\ref{f:jm}.
For a given mass $M$,
there are two possible values of angular momentum $J$,
which correspond to two possible values of radius $R$ in Fig.~\ref{f:krm}.
In the case of NNSs with the BOB EOS,
the branch with the lower $R$ has a larger values of
$J\sim M R^2 f_K$,
because the Kepler frequency $f_K$ increases faster than $R^2$ diminishes on
the Keplerian sequence.
In the case of HNSs, the situation is opposite:
the branch with the lower $R$ has also a lower value of $J$.
Therefore for NNSs the Kepler curve meets the SAI at large $R$,
before it reaches the mathematical maximum of the mass.
This is different from the case of HNSs,
whose curves extend a little further on the unstable branch
after they reach their mathematical maximum, before meeting the SAI,
and thus the maximum mass of the stable configurations coincides
with the mathematical maximum value.
The maximum mass and maximum angular momentum,
as well as the end point given by the SAI constraint,
are obtained with different stellar configurations,
and are labelled by the open squares, triangles, and circles, respectively.
The discussed effects are however very small, of the order of 0.01$\ms$ at most.

In order to visualize better the intricate relations between $M$, $R$, and $f_K$,
we present in Fig.~\ref{f:rhomf}
the mass-radius relations of NS with EOS BOB (upper panel) and DS2 (lower panel)
at various fixed rotation frequencies (dash-dotted olive curves).
The stable configurations are constrained by the Kepler and SAI conditions
at large and small radius, respectively.
At a low frequency ($f=796\,$Hz for HNSs),
the lower boundary of $M$
is fixed by the Kepler condition and the upper boundary by the SAI condition.
As the frequency increases ($f=1082\,$Hz),
the SAI mark point moves to the left side of the mathematical maximum (MaM),
and the upper boundary of $M$ is now fixed by the MaM,
but not anymore by the SAI condition.
This is indicated by the dotted blue curve that passes through the MaMs
for fixed frequency.
As the frequency increases further ($f=1194/1273\,$Hz),
the lower (upper) boundary values of $M$ are fixed by the SAI (MaM/Kepler)
conditions.
Finally, at the maximum frequency the Kepler and SAI conditions
meet at the same point.

\begin{figure}[t]%..............................................................
\vspace{-23mm}
\centerline{\includegraphics[scale=0.31]{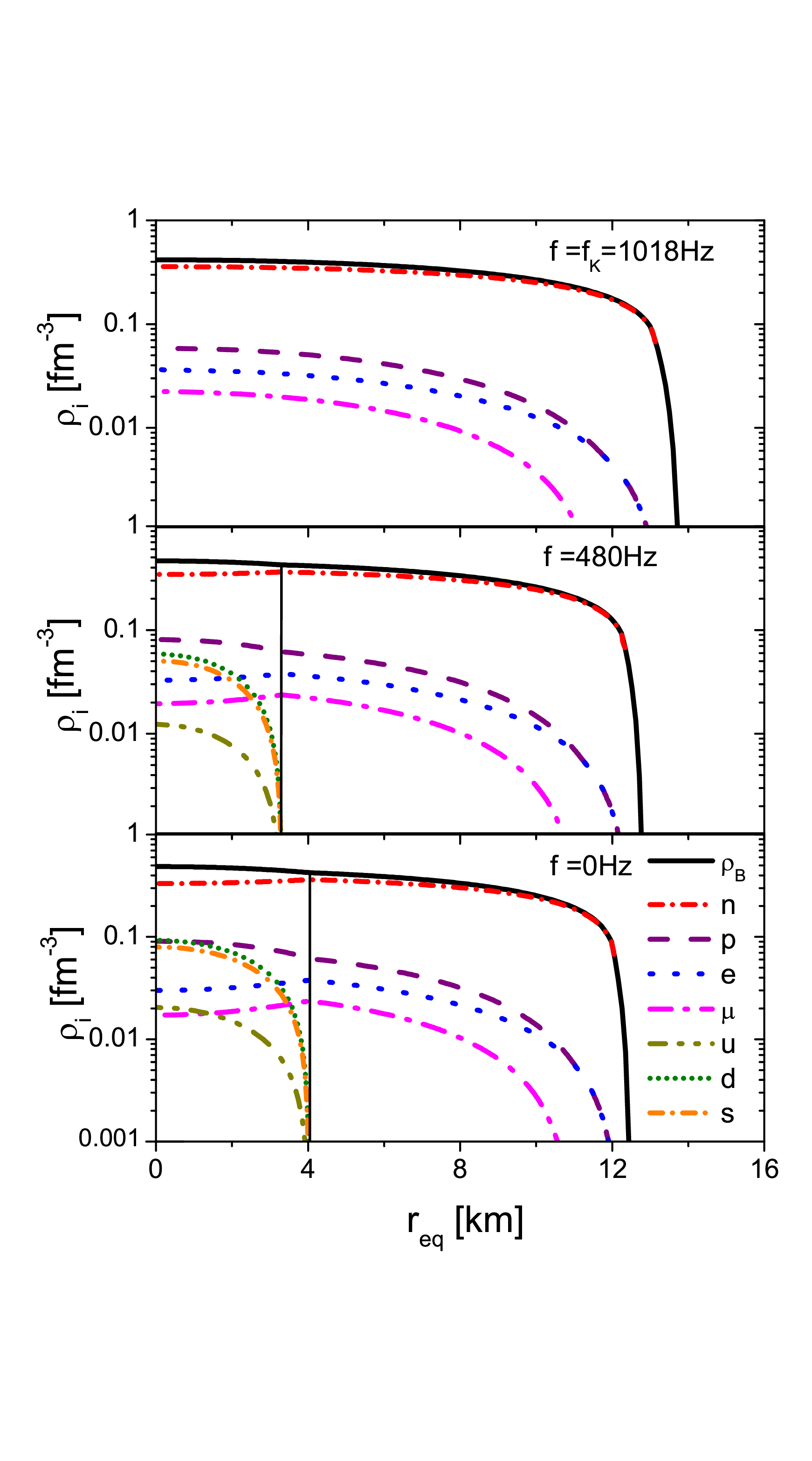}}
\vspace{-26mm}
\caption{(Color online)
Equatorial profiles of particle number densities of a rotating NS
of baryonic mass $M_B= 2.0 \ms$ at various rotation frequencies
with the DS2 EOS.
The vertical solid lines represent the interface of the two phases.}
\label{f:popu}
\end{figure}%...................................................................

\begin{figure}%.................................................................
\vspace{-5mm}
\centerline{\hspace{-6mm}\includegraphics[scale=0.37]{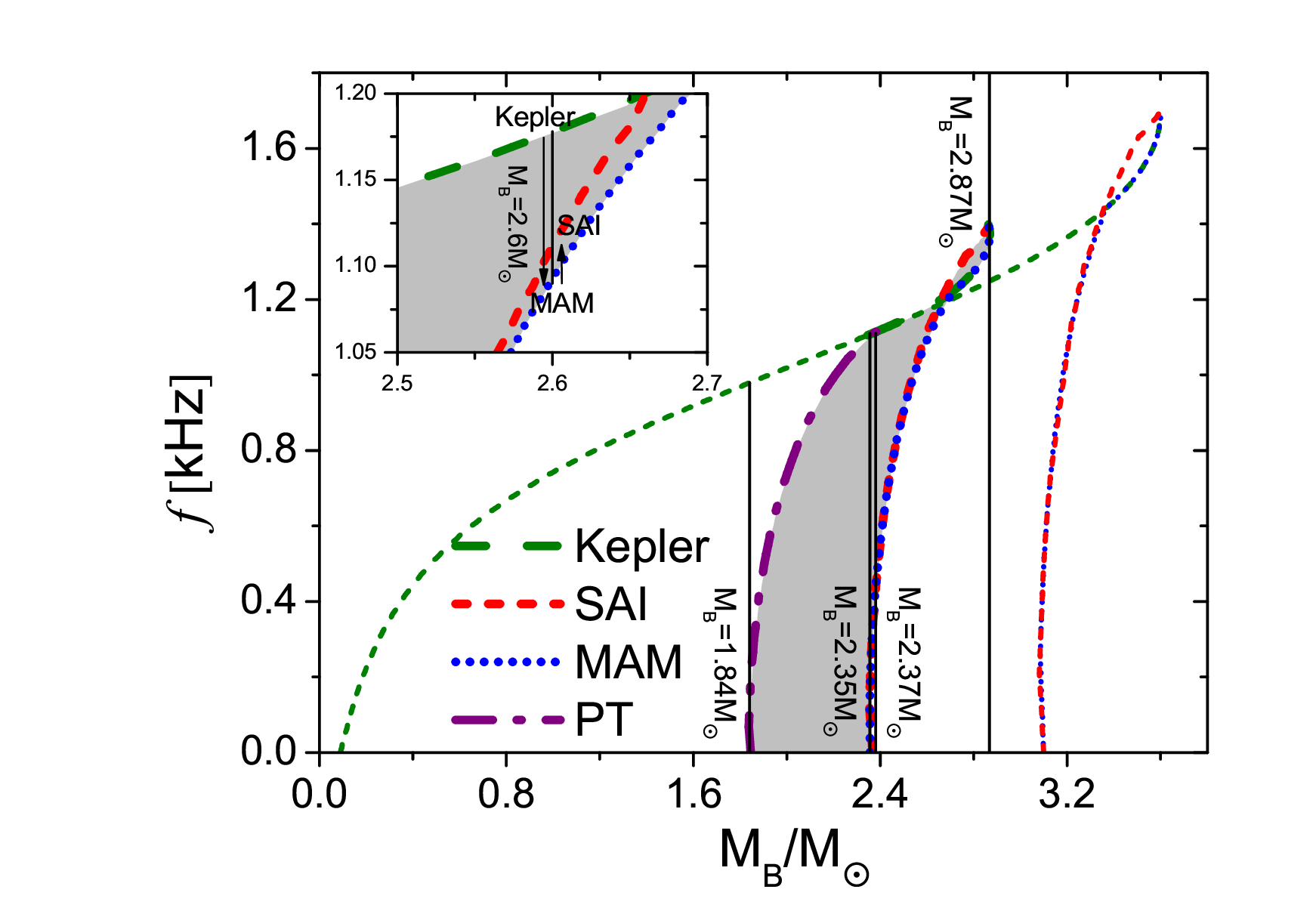}}
\vspace{-5mm}
\caption{(Color online)
The allowed domain of HNSs with the EOS DS2
in the $f$--$M_B$ plane.
The legend is as in Fig.~\ref{f:frm}.}
\label{f:mb}
\end{figure}%...................................................................

\begin{figure}[t]%..................... ........................................
\vspace{-11mm}
\includegraphics[scale=0.30]{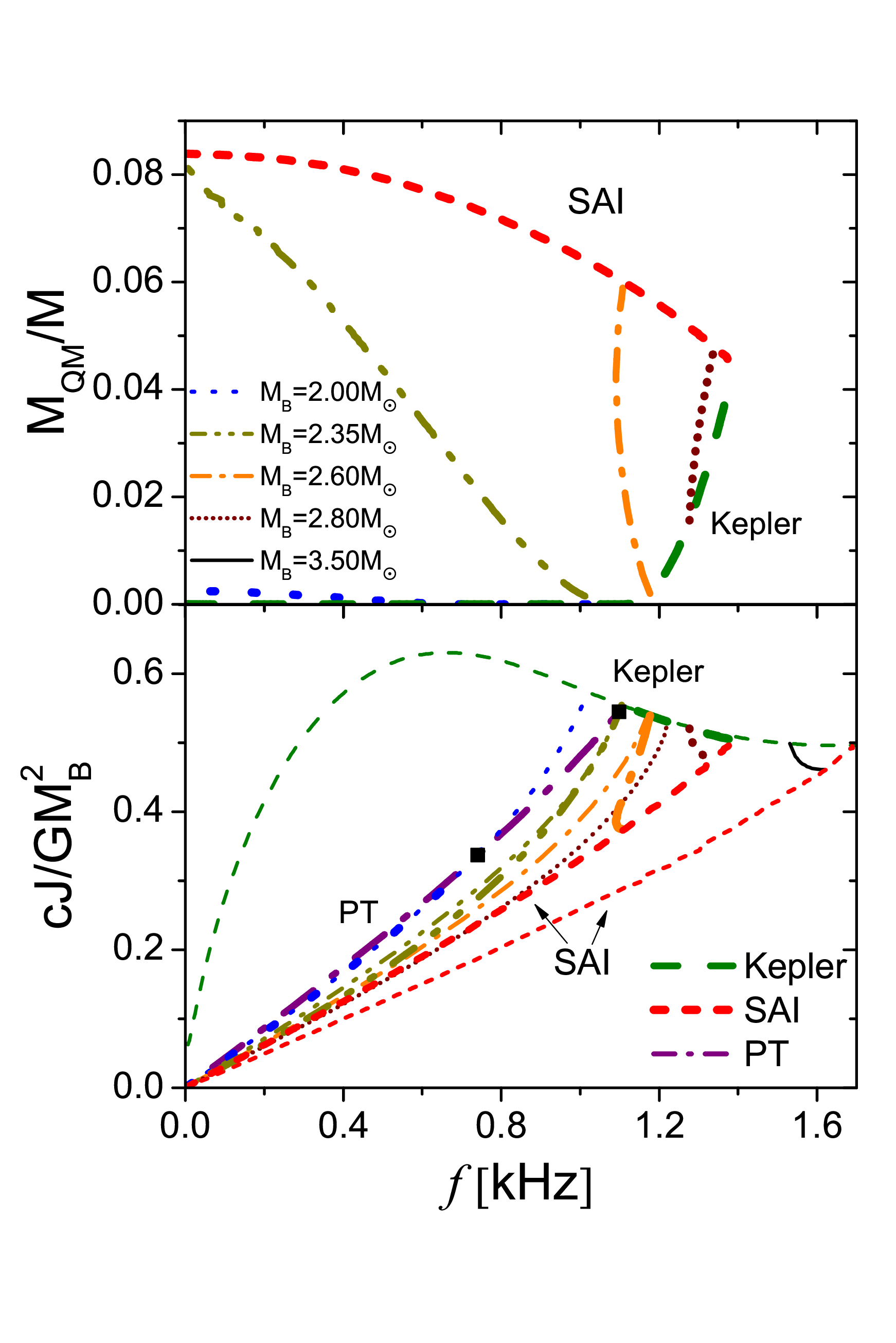}
\vspace{-19mm}
\caption{(Color online)
Mass fraction of QM (upper panel)
and angular momentum (lower panel)
as a function of rotation frequency
for several fixed values of $M_B$.
Bold curves are for HNSs with the DS2 EOS
and thin curves (in the lower plot) represent the results for NNSs.
The markers indicate the onset of the HQ phase transition.
The Kepler, SAI, and PT lines are shown,
as in Figs.~\ref{f:frm} and \ref{f:mb}.}
\label{f:fj}
\end{figure}%...................................................................

In Fig.~\ref{f:frm} we present the allowed domain of NNSs
and HNSs in the $R_\text{eq}$--$f$ plane (upper panel)
and the \hbox{$M$--$f$} plane (lower panel),
together with some observational data.
We use the same conventions as in Fig.~\ref{f:rhomf}, i.e.,
dotted blue curves, dashed green curves, and short-dashed red curves
represent MaM, mass-shedding, and SAI limits.
The allowed region of HNSs with the DS2 EOS
is the grey area delimited by the dash-dotted lila curve (PT),
which represents the onset of the phase transition.
One interesting feature we should mention here is that at high rotation frequency
the mass range is small,
while the range of radii is still large,
corresponding to a flat top of the $M(R)$ curves in Fig.~\ref{f:krm}.
This means the radii are very sensitive to the mass at high rotating frequency.

As discussed above,
the current observations on pulsar masses constrain our parameter to $\alpha<2$,
hence we present the results of HNSs with the EOS DS2.
For smaller $\alpha$ the corresponding (shaded) area of HNSs
will shrink and move towards the lower (upper) boundary of NNSs
in the upper (lower) panel.
The minimum (maximum) mass of HNSs with EOS DS2 is 1.68 (2.02)$\;\ms$
in the static sequence,
and increases as the rotation frequency increases,
while the range concentrates to a single value 2.47$\;\ms$
at the maximum frequency $f=1.4\;\text{kHz}$.
Therefore, in the lower panel of Fig.~\ref{f:frm},
the three stars with lower masses should be conventional NS,
and the others could be HNSs in our DS2 model.

The observational data of the radius still suffer large uncertainties.
In the upper panel we include the sources 4U1820-30 and SAXJ1808.4-3658,
whose mass, radius, and spin are available.
One can see that according to their small radii both sources
should preferably be high-mass compact HNSs in our model,
whereas their masses in the lower panel identify them as preferably
``low-mass'' NNSs.
This can also be seen in Fig.~\ref{f:krm},
where the same data points are reproduced.
However, within the large error bars, both data
are still consistent with our model.
We expect more accurate observations to constrain our parameters
or rule out the model.

%-------------------------------------------------------------------------------
\subsection{Phase transition caused by rotational evolution}

The possibility of a phase transition to QM caused by rotational evolution
has been widely discussed in literature
\cite{Weber1997,Spyrou2002,Ayvazyan2013,Haensel2016}.
For a constant baryonic mass,
a rotating star loses its rotation energy by magnetic dipole radiation,
which makes the star spin down and the central density increase.
When the central density of a NNS reaches a critical value,
the phase transition from hadronic matter to QM will take place,
and the star converts to a HNS.
As the star continues spinning down and the central density continues increasing,
more and more QM appears in the core of the HNS.

This is clearly shown in Fig.~\ref{f:popu},
where we display the change of the number density of all particle species
with rotational frequency in the interior of a star with baryonic mass
$M_B=2.0\ms$ for the DS2 EOS
(corresponding to $M=1.74\ms$ in the static sequence
and $M=1.80\ms$ at the Kepler frequency $f_K=1018$Hz,
see the lower panel of Fig.~\ref{f:rhomf}).
One notes that this star at Keplerian frequency has no QM core,
but as it spins down,
it is compressed to a smaller volume,
which enhances the central density,
and the star is converted into a HNS.
As the frequency decreases further,
the QM mixed phase extends outward from the core
and the region occupied by the pure hadron phase gets narrower.
At the same time, the radius of the star is decreasing.

In Fig.~\ref{f:mb} we present the stellar models with DS2 EOS
in the $f$--$M_B$ plane,
where the same labels as in Fig.~\ref{f:frm} are used, i.e.,
the dash-dotted lila curve represents the onset of conversion
from a NNS to a HNS.
It can be seen that the conversion is possible only in the baryon mass range
$1.84<M_B/\ms<2.37$.
Examples could be the pulsars J1903+0327 and 4U1820-30,
located at the edge of the phase transition boundary in Fig.~\ref{f:frm}.
Above that range, even the fastest rotating stars are already HNSs.
In addition, when the star's baryonic mass is larger than 2.35$\ms$,
the static configuration is unstable,
and the star will collapse to a black hole as it loses angular momentum
and meets the SAI borderline (dashed red curve).
These are {\em supramassive} stars \cite{max}
that will be discussed in more detail in the following.
The maximum baryonic mass for the DS2 EOS is 2.87$\ms$.
The various limits are indicated by vertical lines in Fig.~\ref{f:mb}.

For further illustration,
we show in the upper panel of Fig.~\ref{f:fj}
the fraction of QM in HNSs as function
of the rotation frequency for several choices of fixed baryonic mass
with the DS2 EOS.
The trajectories in the $M$--$R_\text{eq}$ plane
for the same values of $M_B$ are reported in Fig.~\ref{f:rhomf}.
Usually the QM fraction increases with decreasing frequency due to the
increasing density and extension of the QM domain in the star,
see Fig.~\ref{f:popu}.
The maximum value of $8.39\%$ is reached for
the heaviest possible static NS with $M_B=2.35\ms$,
see Fig.~\ref{f:mb}.
This value can be increased by choosing larger values of $\al$ in the DSM,
but then the maximum HNS mass falls below two solar masses.
Supramassive HNSs ($M_B>2.35\ms$) have no static limit
and collapse when reaching the (dashed red) SAI line.
Their QM fraction remains below the maximum static value.

In the lower panel of Fig.~\ref{f:fj}
we show the angular momentum as a function of rotation frequency
for NNSs and HNSs.
The conversion points between NNSs and HNSs
on the PT line are indicated by markers in some cases.
Normal HNSs ($M_B<2.35\ms$) are spinning down
when losing angular momentum in the evolution,
whereas supramassive stars spin up close to the collapse \cite{Cook1994}.
A similar backbending phenomenon is often related to the onset
of the phase transition from hadronic matter to QM
\cite{Weber1997,Glendenning1997,Haensel2016},
but here it occurs for both HNSs and NNSs in supramassive configurations,
in the case of NNSs for $3.10<M_B/\ms<3.59$,
see Fig.~\ref{f:mb}.

In more detail,
for example for the $M_B=2.6\,\ms$ trajectory
in Fig.~\ref{f:fj},
Fig.~\ref{f:rhomf},
and in the inset of Fig.~\ref{f:mb},
the HNS spins down until it reaches the minimum of the
fixed rotation frequency curve ($f=1082$ Hz).
Then it spins up until the final SAI point.
In fact, in the evolution
the maximum angular momentum is given at the Kepler sequence
and the minimum angular momentum at the static sequence or the SAI line.
Therefore, if the lower boundary of the frequency in Fig.~\ref{f:mb}
is not at the static sequence or the SAI line,
there must be a spinup with loss of angular momentum.

Quantitatively, the difference of angular momentum
between NNSs and HNSs with equal baryonic mass
is slight at lower baryon mass ($M_B<2.35\,\ms$),
but becomes important for larger masses,
where the QM content increases
and only HNSs exhibit the spinup phenomenon.

%-------------------------------------------------------------------------------
\section{Conclusion}

We have investigated the properties of rotating HNSs,
employing an EOS constructed with the BHF approach for nucleonic matter
and the DSM for QM,
and assuming the phase transition under the Gibbs construction.
We computed the properties of HNSs in the Keplerian sequence,
respecting the SAI constraint.
HNSs are more compact and
have lower maximum masses and maximum Kepler frequencies than NNSs.
Our results for the maximum mass,
maximum rotation frequency,
and the equatorial radius range
fulfill the current constraints by observational data
of the fastest rotating pulsars.

We also investigated the phase transition induced by the spindown of pulsars
with a constant baryonic mass.
We showed the variation of the QM content under rotational evolution,
and found that the QM ratios are small,
with the maximum value about $8\%$,
in order to respect the current two-solar-mass lower limit of the maximum mass.
We also found that in our model the spinup (backbending) phenomenon
is not related to the phase transition,
but happens in supramassive stars before they collapse to black holes,
which is possible in a narrow range of large mass for both HNSs and NNSs.

\section{Acknowledgments}

We acknowledge financial support from the NSFC (11305144,11475149,11303023).
Partial support comes from ``NewCompStar," COST Action MP1304.

%-------------------------------------------------------------------------------

\end{document}